# Taranga DR1: Analysis of TESS Short Cadence data for years 1 and 2

Sangeetha Nandakumar*[1] | Mauro Barbieri[2] | Jeremy Tregloan-Reed[1]

[1]Instituto de Investigación en Astronomía y Ciencias Planetarias, Universidad de Atacama, Copayapu 485, Copiapó, Atacama, Chile

[2]Terma GmbH, Europa-Arkaden II, Bratustrasse 7, 64293 Darmstadt, Germany

**Correspondence**
*Sangeetha Nandakumar. Email: sangeetha.nandakumar@postgrados.uda.cl

TESS (Transiting Exoplanet Survey Satellite) was launched in 2018 with the purpose of observing bright stars in the solar neighbourhood to search for transiting exoplanets. After the completion of the two year nominal mission, TESS has provided 2 minute cadence photometry of over 200 000 stars. This large collection of light curves opens the possibility to study the statistical and temporal properties of this ensemble of stars.

Most of the currently available data pipelines are designed to work on single sector at a time. We present a new TESS data pipeline called `Taranga`, with the purpose of merging multi-sector light curves, whilst performing a period search for all the observed stars, and stores the statistical results in a database.

`Taranga` pipeline has three components which 1) processes the PDCSAP fluxes of each sector and creates merged PDCSAP light curve, 2) performs a similar operation on the SAP fluxes, and 3) generates the periodograms of the merged SAP and PDCSAP light curves while performing peak identification.

For all the 232 122 stars observed in short cadence in the nominal TESS mission, we provide the merged PDCSAP and SAP light-curves along with their periodograms. We provide a database that has the statistics of all the results produced from `Taranga` of these stars.

**KEYWORDS:**
Methods: data analysis – Surveys – Catalogs – Astronomical data bases : miscellaneous – asteroseismology – Stars: variables: general –

## 1 | INTRODUCTION

Through the contribution from several dedicated space missions like MOST (Matthews et al., 1999), CoRoT (Baglin & Chaintreuil, 2006; Baglin, Michel, & Noels, 2013; Baglin, Vauclair, & COROT Team, 2000; Rouan et al., 1999), Kepler (Borucki et al., 2010; Koch et al., 1998), TESS (Ricker et al., 2015), Cheops (Benz et al., 2020, 2021), the photometric time series data of hundreds of thousands of stars have been made available. The differences in telescope diameter, field of view, angular resolution, length of observations of these missions are varied passing from missions that are dedicated to single stellar targets to all sky observations.

In this context TESS is the first space mission that is able to cover almost the entire sky, providing photometry for a very large sample of bright stars. In the two year nominal mission, TESS has observed over 200 000 stars in both hemispheres with a cadence of 2 minute. These observations are used in both the discovery of exoplanets and for other astrophysical studies for instance, asteroseismology and variable stars.

Light curves and periodograms are the two basic data products in exoplanet transit search surveys. The light curves of the parent stars are used for searching for periodic dips coming from transits of planets, and for understanding their intrinsic variability. Periodograms, Fourier transforms, and phase folding techniques (e.g. Box-Least-Squares Kovács, Zucker, and Mazeh 2002), are used to identify orbital periods, rotational periods, or any other kind of repeated variability in the light curves. The frequency analysis of the periodograms and fourier transforms also enables better understanding of the internal composition of the stars using the knowledge coming from asteroseismology (for reviews, refer to Chaplin and Miglio 2013 and Di Mauro 2016).

With the large abundance of photometric time series coming from space missions, there is a necessity for pipelines that can process the data and extract relevant information that can be used for future scientific studies. The raw data obtained by TESS are processed by the Science Processing Operations Center (SPOC) pipeline (Jenkins et al., 2016) that create the TESS data products available at MAST[1].

In addition to the SPOC processed data products, the Full Frame Images (FFI) with a cadence of 30 minute (10 minute cadence in the extended mission), are publicly released and several independent pipelines were developed for extracting and processing light curves also from these data. Among them, Lightkurve (Barentsen & Lightkurve Collaboration, 2020) creates light curves from Target Pixel Files (TPF) which are simply cutouts around a source found in the FFI, whilst accessing the SPOC processed light curves. The Quick Look Pipeline (QLP) (Fausnaugh, Burke, Ricker, & Vanderspek, 2020; Huang et al., 2020a, 2020b) along with Eleanor (Feinstein et al., 2019) produce light curves from the FFIs. While, finally TESS Asteroseismic Science Operations Center TASOC (Handberg et al., 2021) performs photometry on the TPFs and FFIs to create light curves aimed at asteroseismological studies.

After the completion of two years of the TESS mission, there is an opportunity to create complete, merged light curves of all the stars observed in 2 minutes cadence in this period. Most of the available pipelines focus on processing the light curves sector by sector, and therefore, there is no single repository of merged PDCSAP and SAP light curves of TESS. There is also no database which has the statistics of the complete light curves of TESS and their periodograms. However, the merged SAP and PDCSAP light curves are important in studying the astrophysical signals in stars. The merged SAP and PDCSAP light curves from the TESS will have stars observed in different lengths and showing different properties. Some of these stars, especially the long period variables, can only be studied in the merged light curves and their periodograms. Although many planets have already been discovered using only the partial data from year 1 or year 2, the long term variations in the transit or the host star can only be studied with the merged light curve.

Many classification and clustering algorithms used in astronomy requires a large sample of light curves and periodograms. Additionally, as the different type of stars have different statistical properties, the statistics of the merged light curves of a large number of stars can help in identifying interesting stars. Due to the lack of available merged light curves, these studies have to be done on a case by case basis instead of on a larger scale. A repository of merged light curves and periodograms saves time in data collection and preparation. With LightKurve (Barentsen & Lightkurve Collaboration, 2020), it is possible to create a single light curve of a target observed in multiple sectors. LightKurve creates light curves of each sector after downloading all the TPF data and stitches them together to produce a single multi-sector light curve. However, this has to be done for each target individually. This approach is time consuming both in the development of the software as well as running it on a large sample of stars. We have developed a pipeline with an intention to minimize this limitation. We have run our pipeline on the stars observed by TESS in the first two years and created data products (light curves, periodograms, and their statistics). The results of our pipeline will serve as an addition to the already existing data products from other pipelines.

The article is organized as following: Sec 2 describes the TESS observation strategy, the type of data products from TESS, and the kinds of light curves (PDCSAP and SAP). Sec 3 introduces the pipeline and gives a brief description of the components of the pipeline. Sec 4, Sec 5, and Sec 6 describes the procedures followed for processing of PDCSAP, SAP, and generating the periodograms. Sec 7 focuses on the preliminary results and on the discussion of the results and in Sec 8, we give our concluding remarks.

## 2 | OBSERVATION AND DATA

TESS satellite is equipped with four wide field cameras each with a field of view of 24° × 24°. The cameras are arranged in a line in order to observe 24° × 96° of the sky at any given time, providing a pixel scale of 21" per pixel.

TESS is in a stable, elliptical, and 2:1 lunar resonant orbit chosen to provide a long observation period suitable for detecting exoplanets while also protecting the satellite from Earth's radiation belt. From this orbit, TESS covers nearly all-sky with four cameras. It observes a part of the sky (called Sector) for two orbits (approximately 27.4 days) with camera number 4 fixed on the ecliptic pole. At the end of each orbit, observations are stopped for approximately 16 hours for data downlink to the ground stations. After the completion of one sector, the FOV moves eastward around the ecliptic pole by 27° to observe the following sector. Each sector will have two gaps in observation after the two orbits during the data downlink. There is significant overlap between sectors especially near the ecliptic poles. Therefore, some stars are observed in more than one sector leading to nearly one year of observation times. TESS covered the southern and northern hemisphere stars in sectors 1 to 13 and sectors 14 to 26 respectively. Some sectors were shifted upwards due to stray light from the solar system bodies which created a different footprint from year 2 with respect to year 1 as portrayed in Fig1 .

---

[1] https://archive.stsci.edu/missions-and-data/tess

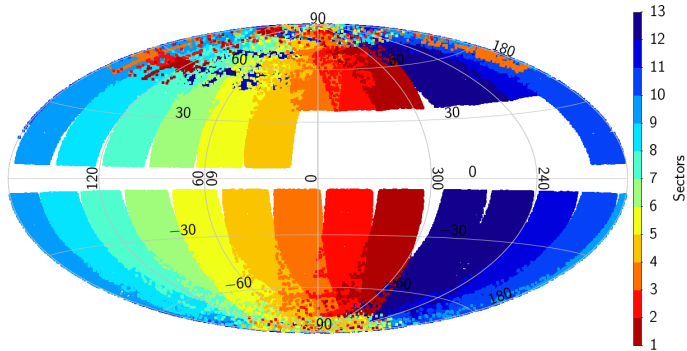

**FIGURE 1** Footprint of TESS observations for the nominal mission in ecliptic coordinates. The color scale indicate the Sector observed, the numbers start from 1 from each hemisphere.

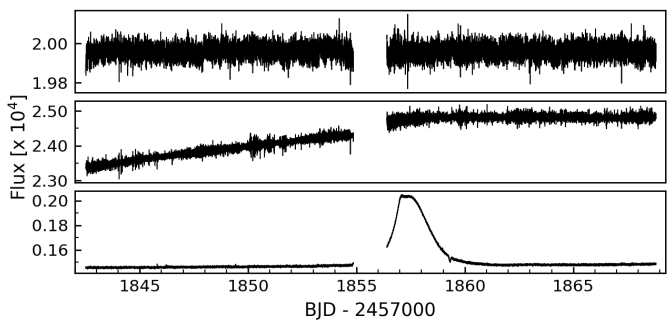

**FIGURE 2** Example light-curves for TIC 159336506 (BD+83 450) a G0 subgiant. The horizontal axis is in BJD, the vertical axis is in e$^-$/s. In the top panel : is presented the PDCSAP flux, in the middle the SAP flux, while in the bottom the background BKG.

For the TESS nominal mission there are two main kind of photometric data: the Full Frame Images (FFI) and the Target Pixel Files (TPF)

The FFIs are the images obtained by each camera with an equivalent exposure time of 30 minutes. The TPFs can be for both 30 minute cadence and 2 minute cadence data. For a limited collection of stars (about 20 thousand per each sector) selected based on their potential for hosting exoplanets or asteroseismology (Stassun et al., 2018), TESS provides TPF with 2 minute cadence. In the following we call these files Short Cadence (SC) files. The TPF targets with 2 minute cadence are processed by the SPOC pipeline (Jenkins et al., 2016) to produce the SC light-curve files. The SPOC pipeline produce one SC file per star per every sector, so does not create stitched light curves for stars observed in more than one sector.

Each SC file contains three types of photometric data as function of time: SAP (Simple Aperture Photometry) flux, PDCSAP (Pre-search Data Conditioning SAP) flux, and the background fluxes shown in Fig. 2 . SAP (Morris et al., 2017) fluxes are extracted from an aperture selected for a target in a way to optimize the S/N and then corrected for background. Although it is corrected for background, other systematics, such as the ones induced by the rolling of satellite, still exist. The PDCSAP light curve is created by removing such systematics from the SAP by making use of Co-trending Basis Vectors (CBV) and it is created for each sector a star is observed.

The background flux is contained in the SC files and it is estimated by the SPOC pipeline at the location of the target star pixel by fitting a two-dimensional surface to the background pixels in a procedure similar to the one used by Kepler (Jenkins et al., 2010). The date of the SC files is in Barycentric Julian Date in TDB scale.

In the following we will use term *original* for referring to the light-curves obtained from the MAST archive, we apply this term also to the components SAP, PDCSAP and BKG.

## 2.1 | Data

From the MAST archive we retrieved all the SC light-curves of year 1 and 2. There are over 400 000 observations from the first two years of TESS mission. Each sector has approximately 20 000 targeted observations. The number of SC files per sector is given in Table 1 .

Not every targeted observation correspond to a unique star. We need to keep in mind that there is a significant overlap between sectors. A single star can be observed between 1 and 13 times, and there are $2^{13}$ possible combination of sector observations. The number of possible combinations for one single stars depends on its proximity to the TESS continuous viewing zone (i.e. how near it is from the ecliptic pole).

A total of 2023 different combinations of sectors were used, of which 538 combinations were used both in years 1 and 2, 544 were used only in year 1, while 941 were solely used during year 2. The most frequent combinations in year 1 is sector 13 with 8249 stars observed, for year 2 is sector 17 with 6693 stars. There are 1828 and 1864 stars observed in all thirteen sectors of both year 1 and 2.

| Year 1 | | Year 2 | |
|---|---|---|---|
| Sector | $N_t$ | Sector | $N_t$ |
| 01 | 15 889 | 14 | 19 993 |
| 02 | 15 994 | 15 | 19 996 |
| 03 | 15 993 | 16 | 19 999 |
| 04 | 19 997 | 17 | 19 996 |
| 05 | 19 996 | 18 | 19 997 |
| 06 | 19 995 | 19 | 19 997 |
| 07 | 19 995 | 20 | 19 997 |
| 08 | 19 994 | 21 | 19 995 |
| 09 | 19 996 | 22 | 19 999 |
| 10 | 19 999 | 23 | 18 289 |
| 11 | 19 990 | 24 | 19 997 |
| 12 | 19 990 | 25 | 19 993 |
| 13 | 19 997 | 26 | 19 993 |

**TABLE 1** Number of targeted observations in each sector. The total number of observations (i.e. light curves) is 427 794 observations in the two years of the TESS nominal mission.

| $S_T$ | $N_s$ | P [%] |
|---|---|---|
| 1 | 152 677 | 65.9 |
| 2 | 38 979 | 16.8 |
| 3 | 13 017 | 5.6 |
| 4 | 5 372 | 2.3 |
| 5 | 2 765 | 1.2 |
| 6 | 2 013 | 0.9 |
| 7 | 1 473 | 0.6 |
| 8 | 1 217 | 0.5 |
| 9 | 1 509 | 0.6 |
| 10 | 1 588 | 0.7 |
| 11 | 2 678 | 1.1 |
| 12 | 5 171 | 2.2 |
| 13 | 3 682 | 1.6 |
| total year 1 | 128 214 | 55 |
| total year 2 | 103 908 | 45 |

**TABLE 2** Number and percentage of stars observed only in one or more sectors for both year 1 and 2. $S_T$ : total number of sectors in which one stars was observed $N_s$ : number of stars P : percentage (%) of stars to the total observed.

Considering only the unique stars observed in the first two years of operation, TESS provides photometry of a substantial number of 232 122 stars. Table 2 shows how many stars have been observed in one or more sectors. Nearly 65% of the total number of the unique stars are observed only in single sector signifying that the majority of the SC files correspond to a unique star. The other significant portion of observations (16%) is occupied by stars observed in two sectors.

## 3 | TARANGA PIPELINE

We have developed our own pipeline for analyzing TESS Short Cadence data and creating merged light-curves from the different sectors, and performing statistical analysis and signal detection on the light-curves. The goal of this pipeline is for the creation of processed light curves and periodograms that can be used for exoplanet studies and asteroseismology. The second goal was also to have a pipeline that could merge different kind of light-curves coming from different survey or satellites.

We call our pipeline `Taranga` from Kannada language: 'Taranga' means 'wave' in Kannada language, while 'Tara' translates to 'star'. Hence `Taranga` could be considered a crasis of both words with the meaning 'star waves'.

`Taranga` is composed of three parts: the PDCSAP processor (PDCP), the SAP merger (SAPM), and the periodogram generator (GPER).

The workflow of the pipeline is given in Fig. 3 . The pipeline begins with user input of the TICID of a target star, from the file SectorInfo which controls how many times the star was observed and in which path its files are stored. Independently of the number of the observations the pipeline will continue through the processing of the PDCSAP fluxes (PDCP), and the calculation of the related periodograms (GPER). If the star is observed in more than one sector it will processed also by the SAPM and the resulting light-curves will used for calculating the periodograms. The results of all these steps are stored in different files in forms of light-curve, periodogram and summary table.

- PDCP processes the SC observations for creating a normalized PDCSAP light curve. If the star is observed in multiple sectors, a merged light curve is created combining the fluxes of the observations from all the sectors.

- SAPM operates solely on the SAP fluxes of stars observed in two or more sectors: it creates a merged and reconnected SAP light-curve.

- GPER creates Lomb-Scargle periodogram (VanderPlas, 2018) and performs frequency identification in the periodogram, it works on the output coming from PDCP or SAPM.

The pipeline is run separately on year 1 and year 2 stars as there is no overlap between the regions of the sky observed in the two years. Therefore, the stars in the TICID catalog are separated according to the year in which they are observed. The SectorInfo catalog is prepared separately for the two years of observations.

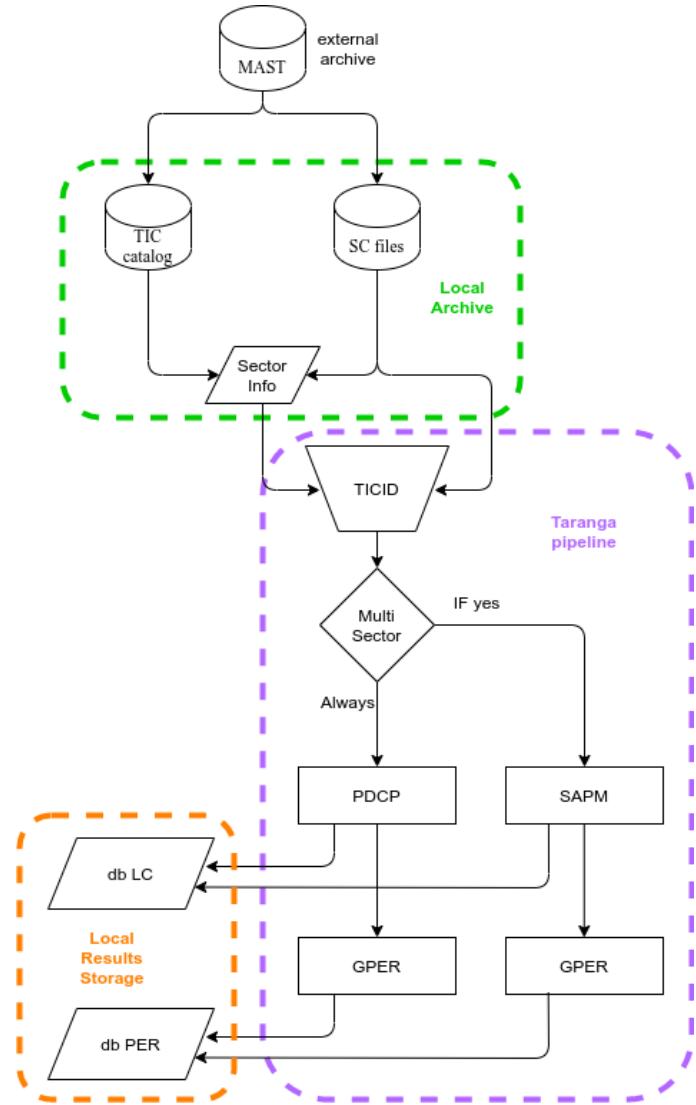

**FIGURE 3** Workflow of the `Taranga` pipeline. The green block is the local archive of input data composed by the SC light-curves and the TIC catalog downloaded from MAST archive. The SectorInfo file is an hash table that allow to identify the path of each file corresponding to a particular star. This table is created from the TIC catalog and the list of the SC files. The purple block is the `Taranga` pipeline with all its components, while the orange block is the ensemble of output files from `Taranga`.

## 4 | PDCP: PROCESSING THE PDCSAP FLUXES

The merging and normalizing of the PDCSAP light curves have to be done by taking into consideration the various effects introduced by the observation strategy. One of the most commonly observed effect is the sudden change in the median flux levels known as "jumps" (see Fig. 4 ). Jumps are expected when there is an attitude tweak of the satellite or pointing of the satellite. In the case of TESS, such jumps can be expected after the change in pointing at the end of downlink or the rolling of the satellite from one sector to another due to the different position of the star on the CCD, or the change of CCD in which the star falls (García et al., 2011; Handberg & Lund, 2014; Jenkins et al., 2010; Lund, Handberg, Kjeldsen, Chaplin, & Christensen-Dalsgaard, 2017). The jumps are additive in nature and, often, an additive correction is performed to process this effect (García et al., 2011)

The purpose of the PDCP procedure is to merge the PDCSAP fluxes of the different sectors and normalizing them. The merging needs to take into account the "jumps" between each sector. While merging, we shift the PDCSAP fluxes of each sector so that the the median flux values of all the sectors are brought to the median flux of the first observed sector, that we take as reference sector. The steps used to merge and normalize the PDCSAP fluxes are the following.

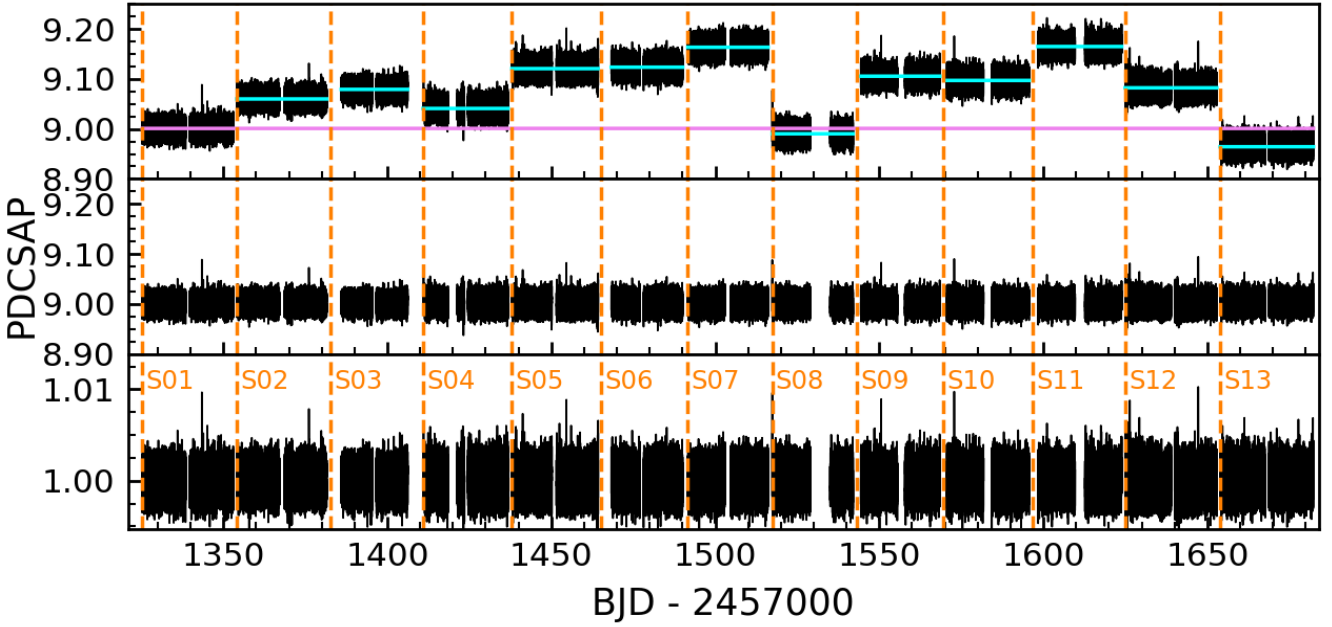

**FIGURE 4** Light-curves of TIC 25063296 (TYC 9156-1430-1), is a late G main sequence star. The star was observed during year 1 in all 13 sectors. All the light-curves are in (in $10^3$ e⁻/s). Top: Original PDCSAP fluxes. The orange dashed verticle dashed lines represent the beginning of observation of the sector. The shift in flux levels can be seen clearly. The cyan horizontal lines are the median PDCSAP flux of each sector while the purple horizontal line is the reference flux value (median PDCSAP flux of the first sector) with respect to which all sectors are shifted. Middle: PDCSAP fluxes after shifting of the fluxes of each sector with respect to the first sector. Bottom: Normalized PDCSAP light curve. The median of the original light curve is 9084 e⁻/s while that of the merged light curve is 9001 e⁻/s which is closer to the median flux value of the first sector (9000 e⁻/s) used as reference.

1. Calculate the median flux ($M_i$) of each sector *i*.

2. Define the median flux of the first observed sector $M_1$ of the star as the reference flux

3. Calculate the difference of each $M_i$ fluxes respect to the reference flux $M_1$: $S_i = M_i - M_1$

4. For each sector subtract $S_i$ to shift the fluxes with respect to the first observed sector

5. Calculate the median flux of the merged light-curve

6. Divide the light-curve by the median flux in order to obtain the normalized light-curve

Sometimes the PDCSAP light-curves contains jumps between the two block of data (first and second orbit) of a single sector. We measured these jumps in all the SC files, calculating the relative difference between the median PDCSAP in the first and the second orbit:

$$\Delta M_O = \frac{M_{O1} - M_{O2}}{M_{O1}} \quad (1)$$

where $M_{O1}$ and $M_{O2}$ are respectively the median of PDCSAP in each orbit of a sector. As expected the distribution of this difference is peaked around zero and it has a very sharp distribution (see Fig. 5). Values of $|\Delta M_O| > 1\%$ correspond to approx 0.9% of the total of the observations (4183 SC files over 479 192). In these cases in order to mitigate the effects of the presence of a step in the normalized PDCSAP, that could inject spurious high frequencies in the periodogram and increasing its background noise, we use the calculated difference $\Delta M_O$ as shift for the median PDCSAP of the second orbit. The stars for which we have applied this correction are flagged in our final database.

All the processes of light-curves merging and normalization is exemplified in Fig. 4. After the creation of the normalized light-curve we then calculate basic statistical figures (standard deviation, skewness, kurtosis) both on the merged PDCSAP and normalized PDCSAP light-curves, and these values are stored in the final database. The two panels of Fig. 6 show two examples after being processed by the PDCP normalization. TIC 278352995 was observed in three sectors during year 2 (S14, S15, and S16). The variations in the minimum and maximum flux within the same sector are not changed but there are no jumps in between two sectors. The correction of jumps is seen better in the example of the eclipsing binary TIC 352165018 which was observed by TESS in the same sectors (S14, S15, and S16). The depths of the primary and secondary eclipses across different sectors are consistent.

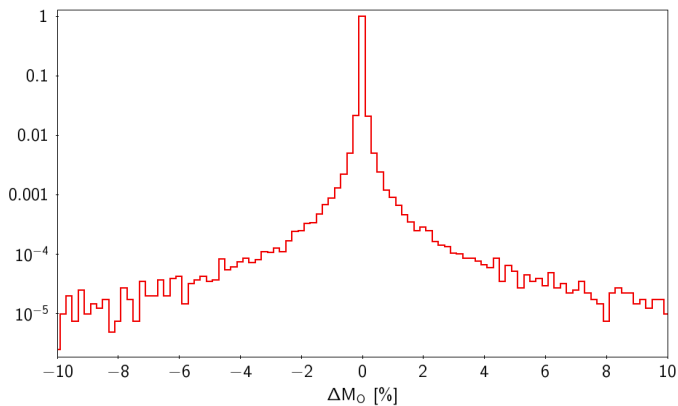

**FIGURE 5** Normalized distribution of $\Delta M_O$. The distribution is peaked at zero with values of quartiles Q01 and Q99 equals to -0.4% and 0.42%.

## 4.1 | Median vs mean PDCSAP for normalization

Both for merging and normalization, the choice of median or mean flux will not affect the final light-curves. This is owing to the method we perform to merge and normalize the light curves. When we make the merging of the PDCSAP, we are assuming that the distribution of fluxes of the reference sector and other sectors are identical except for the addition/subtraction of a constant flux value which we refer to as shift. As a consequence of this assumption, the difference between the mean fluxes of a sector and the reference sector will be the same as the difference between their medians. The only effect of the choice of median or mean on the final merged PDCSAP light curve is the base level of the flux which will be centered around the median or mean flux of the reference sector.

To test whether the choice of the median used for calculating the shift affect the final light curves respect to another measure of the central measure like the mean, we consider all the stars observed in more than one sector. For these stars, we calculate the mean and median of the PDCSAP fluxes of each sector.

If a star is observed in $n$ sectors (where $n > 1$), it will have $n$ mean and median fluxes for PDCSAP, we call them respectively $f_{\mu,i}$ and $f_{M,i}$ where $i$ is the corresponding sector number. For each star we calculate the following quantities $\mu = mean(f_{\mu,i})$ and $M = median(f_{M,i})$. Finally we calculate the percentage difference between $\mu$ and $M$ for all these stars.

The normalized difference $nD = (M - \mu)/M$ of the PDCSAP fluxes (Fig. 7) is peaked at 53 ppm. The difference compared to the order of magnitude of the fluxes is small meaning that the choice of mean or median will not make significant difference in the value of shifts calculated. Additionally, this difference is smaller than the typical TESS noise floor of 60 ppm. Therefore, there will not be significant changes in the final light curves whether we use mean or median.

We consider the same sample of stars to verify that the merging has corrected the "jumps". Let the normalized rms of the original PDCSAP fluxes be $\sigma_{ori}$ and that of the merged PDCSAP be $\sigma_{mer}$. The normalized rms to be an indicator of how well the jumps in the light curves have been corrected. An unprocessed light curve will have "jumps" in the flux measurements between sectors leading to a wider distribution of the fluxes. In contrast, the corrected light curve will have a narrower distribution of the fluxes as all the sectors line up. Therefore, the $\sigma_{ori}$ is expected to be larger than the $\sigma_{mer}$, in Fig. 8 we shows the normalized histogram of the distributions of $\sigma_{ori}$ and $\sigma_{mer}$. As is clearly seen the merging procedure reduce the rms of a factor close to 10, the large queue is however generated by faint stars that have an intrinsic high photon noise.

## 5 | SAPMERGER: MERGING THE SAP LIGHT-CURVES

Retaining signals of length longer than one sector is particularly important for the evolved stars as their rotation or pulsation periods are very long. Therefore, there is a need to have a merged SAP light curve to study astrophysical signals (generally low frequency signals) that may be missed from the PDCSAP light curves. This effect is not unique to TESS but can also be seen in Kepler stars which uses a technique similar to TESS for creating PDCSAP light curves (eg. (Still, Howell, Wood, Cannizzo, & Smale, 2010)).

The SAP fluxes also exhibit jumps between sectors similar to the PDCSAP. For the PDCSAP flux we assumed that there is no trend in the regions of discontinuity which allowed the use of median value to calculate the shift. This assumption is no longer true for the SAP fluxes as there are some increasing or decreasing trends of the SAP fluxes in each sector as seen in Fig. 9, Fig. 10, and Fig. 11. The shifts have to be calculated considering the trend that could be present in the gap. But the trend can be estimated only between two consecutive sectors as it is difficult to determine the trend for longer intervals (which is at least the length of a sector for stars observed in non-consecutive sectors). Therefore, we only consider stars which are observed in at least two consecutive sectors. Table 3 shows the number of stars (with a minimum of two sectors of observation) for which SAP merging is possible. For a significant number of stars ($\approx$ 90% of the stars observed in at least two sectors), SAP merging can be done. We note that Table 2 and Table 3 , cannot be directly compared because columns S and $n_c$ does not represent the same quantity. As an example, a star could be observed in 12 sectors but not consecutive, for example from sector 1 to 6 and from 8 to 13, in this case the largest consecutive number of sector is 6.

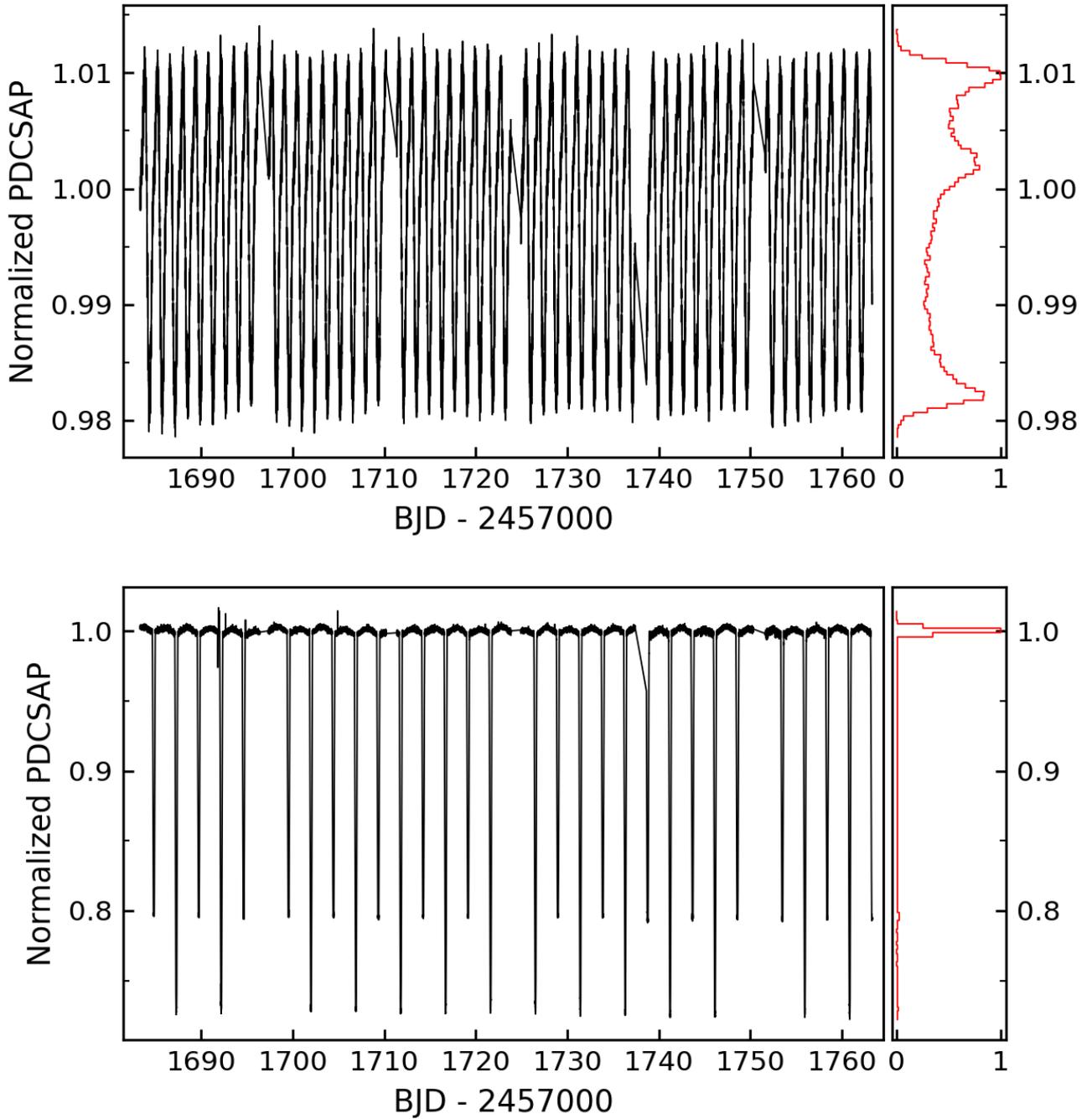

**FIGURE 6** TIC 278352995: BD+48 3140, an A0V star of $\delta$ Scuti type, observed in 3 sectors during year 2 (TOP). TIC 352165018: V2529 Cyg, a F2 an Algol type eclipsing binary (BOTTOM).

For reconnecting the SAP fluxes we developed a simple algorithm that is mostly suited for recovering long period variation on stars like pulsation on AGB star or very slow rotation period on main sequence stars. The algorithm that we implemented is based on the assumption that the stellar variation present in the SAP fluxes are real (i.e. not caused by systematics) and that in a short time scale could be represented by low order polynomials.

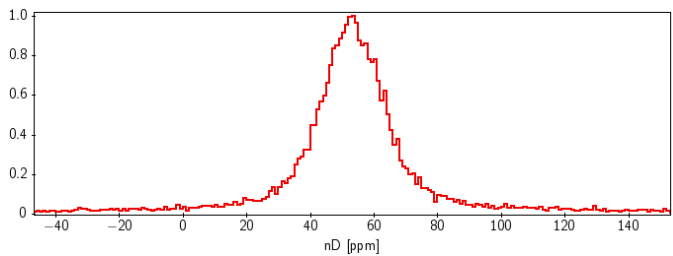

**FIGURE 7** Normalized histogram of the difference between mean and median of the PDCSAP fluxes $nD = (M - \mu)/M$ for stars observed in more than one sector. The

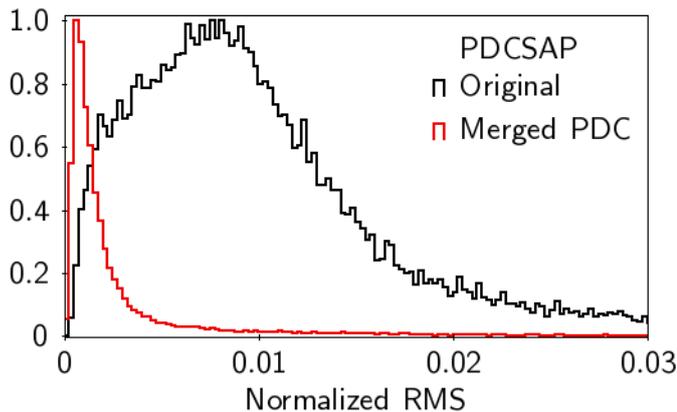

**FIGURE 8** Comparison of the histograms of the normalized rms of the original and merged PDCSAP light curves. The black and red lines show the normalized histogram of $\sigma_O$ and $\sigma_M$ respectively

| $n_c$ | $N_s$ |
|---|---|
| 2 | 40 452 |
| 3 | 10 129 |
| 4 | 3 444 |
| 5 | 2 089 |
| 6 | 1 298 |
| 7 | 1 009 |
| 8 | 1 187 |
| 9 | 1 776 |
| 10 | 1 473 |
| 11 | 4 782 |
| 12 | 523 |
| 13 | 3 682 |

**TABLE 3** Number of stars observed in at least two sectors with one or more connections between sectors for SAP merging. $n_c$ : largest number of consecutive sectors, $N_s$ : number of stars observed in consecutive sectors. A connection is defined between any two consecutive sectors of observations. Out of 79 436 stars observed in at least two sectors, 71 844 stars have a minimum of two consecutive sectors of observations while 7 592 stars have no consecutive sectors available.

Let us consider a star observed in many consecutive sectors (eventually with some missing sectors). We call the first observed sector the reference sector ($\alpha$), and its flux $F_\alpha$, the next sector observed that need to be consecutive to the previous, as sector $\beta$ and its flux $F_\beta$.

1. We fit a $n$-order polynomial to $F_\alpha$ of the last $\Lambda$ days of observation of sector $\alpha$. We extrapolated this fit forward for $\Lambda$ days after the end of the sector ($E_\alpha$).

2. We fit a $n$-order polynomial to $F_\beta$ of the first $\Lambda$ days of observation of sector $\beta$. We extrapolated this fit backward for $\Lambda$ days before the beginning of the sector ($E_\beta$).

3. We calculate the median value of the difference of the two extrapolations: $\Delta E = \text{median}(E_\alpha - E_\beta)$

4. We shift the flux of the sector $\beta$ by the quantity $\Delta E$ obtaining a new flux $F_{\beta,new}$

5. $F_{\beta,new}$ will act as the reference sector to the next consecutive sector if present.

The algorithm is portrayed graphically in Fig. 9 .

The first step of SAP merging is identifying all the stars observed in more than one sector from the SectorInfo catalog 3. Then we check if there is at least one connection between sectors. If there are no connections, then the star is not considered for SAP merging. If there are connections, then the sectors are grouped into blocks.

In presence of a disconnected sector (a sector with no immediate consecutive sector), we proceed iteratively until we find a block of two or more consecutive sectors, and we shift the flux of the disconnected sector respect to the median of the SAP fluxes of the previous reconnected block of sectors.

The values of $n$ and $\Lambda$ were chosen after various experiments on long period variables on the AGB branch The value of $\Lambda$ could not be lowered below the time incurring between two consecutive sectors (up to 1.7 days), and conversely cannot be too large because it could lead to a poor fit of the fluxes. We finally determined that the values of $n = 2$ and $\Lambda = 5$ days provides very good results for studying variation on AGB stars with periods larger than 5 days.

In Fig. 10 and Fig. 11 we show two examples of stars for which SAP reconnection was made. In these cases the SPOC procedure over-corrected the SAP light-curves introducing additional noise in some sectors (first two panel from top). In the panel labelled 'SAP median shifted' we show what could be the output of a simple median shifting on SAP fluxes, and is possible to observe this method is not capable to preserve the original content of the signal. It is interesting to note that in some cases like the one of AX Men, the original SAP light curve is very close to the reconnected one because the star fall on the same CCD in each sector in which was observed granting a constant response.

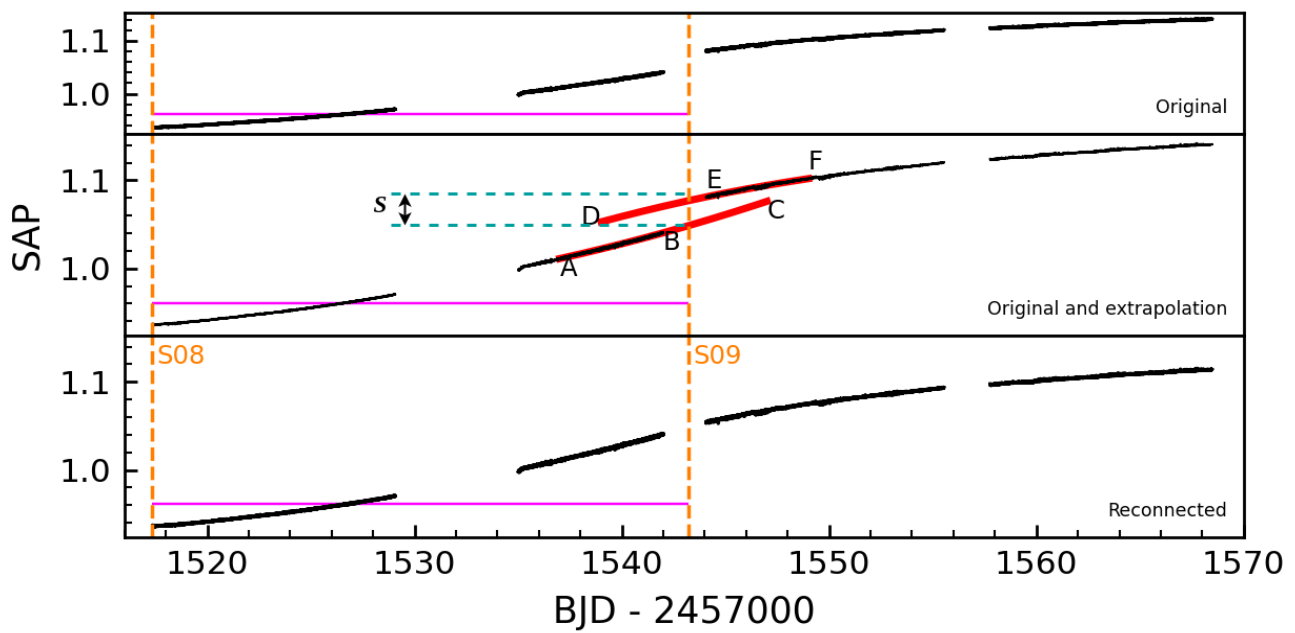

**FIGURE 9** Reconnection of SAP light-curve of TIC 93723838: EP Vel, a Mira type (SpT: M6) observed during 2 sectors in year 1. The orange dashed vertical dashed lines represent the beginning of observation of the sector, and the purple line in all the three panels refer to the median SAP flux of S08. The vertical axes are the SAP fluxes in (in $10^6$ e$^-$/s) Top: The original SAP light-curve. Middle: A second order polynomial (in red) is fitted between the points *A* to *B* and extrapolated between the points *B* to *C*. Similarly on the following sector, a second order polynomial is fitted between the points *E* to *F* and extrapolated between the points *E* to *D*. The letter *S* indicates the shift between the SAP fluxes of the two sectors. Bottom: Reconnected SAP light curve.

We note that the merged SAP light-curve created with this method needs further processing to remove the trends due to the systematics. One of the most common problem on SAP light-curves is the presence of stray-light from the Earth. Its correction require a the use of the CBV files and the simultaneous treatment of many stars in the same regions of the CCD in order to identify the same patterns. The stitching algorithm does not directly take into account the instrumental scatter found at the start and end of the sectors. By setting the length of time to five days, and then extrapolating both forward (end of sector) and backward (start of next sector), we use 10 d of data to generate the extrapolation function, of which we take the median value. This minimises the impact of instrumental scatter, to the sector correction. While it could be argued that it would be better to mask the first and last day of each sector, before performing the extrapolation, any astrophysical processes that could be taking place simultaneously would also be missed, adding an unknown systematic to the proceeding data.

On the other side the star with variations faster than 5 days are not well reconnected by our algorithm. When examining the original and reconnected SAP light curves, we find that the algorithm over/under corrects for stars with a variability of between 4.0 to 4.5 d. This correction happens even when there was no original shift between two sectors. In addition for stars with variability less than 3 d, the algorithm puts all sectors in a single line. If there are gaps of one or more sectors, then it creates a jump-like artifact. Therefore, for targets with variability < 4.5 d, we recommend that the targets are examined on a sector by sector bases, due to $P \ll 27$ d. Another issue that we found is that many stars observed during sector 4 and 5 exhibit a common pattern of increase in their SAP values, and our algorithm reconstruct this variation like a long bump that slowly decrease over many sectors. All the stars that exhibit this behaviour cannot be used for period analysis. In a future release of our data we will address these issues integrating in the algorithm the use of the CBV files.

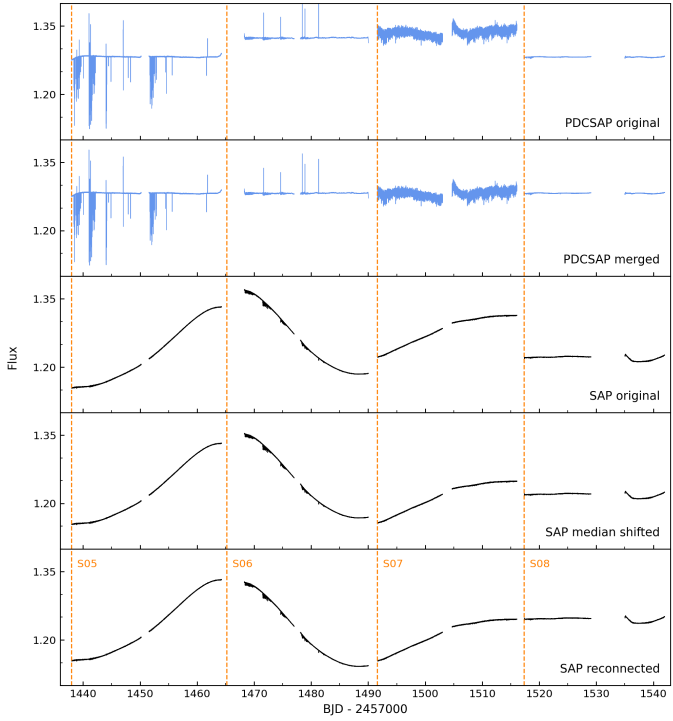
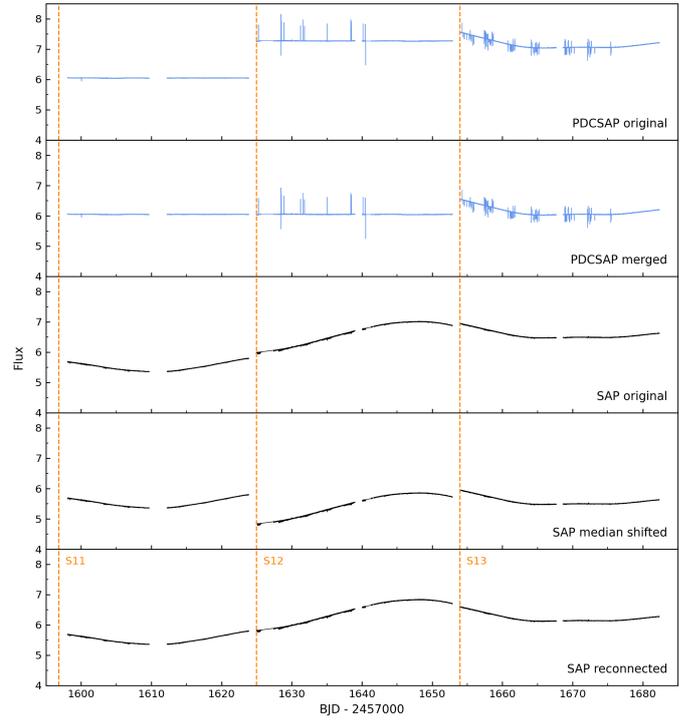

**FIGURE 10** Comparison of different type of light-curves for TIC 255565783: HD 45555 (M2III), a long period variable. (the fluxes are in $10^6 e^-/s$ ) First panel from above: original PDCSAP fluxes. Second: merged PDCSAP fluxes. Third: original SAP fluxes. Fourth: SAP light curve after merging using a technique similar to the one used for PDCSAP. Fifth: SAP light curve after using the reconnection algorithm.

**FIGURE 11** Comparison of different type of light-curves for TIC 278140574: AX Men, a long period variable (the fluxes are in $10^4 e^-/s$). The panels are as in Fig. 10 .

## 6 | GPER: GENERATING PERIODOGRAMS

Periodograms or Fourier analysis are the classical tools for studying time-series in the frequency domains. TESS SC light-curves have many missing data due to the interruptions in the observations due to the passage of the satellite at perigee, and some sector could be missing, hence we calculated Lomb-Scargle periodograms of TESS SC data since they are more suited for frequency analysis of time-series with missing data.

The minimum frequency $f_{\min}$ depends on the length of observation time for each star which can range approximately between 28 days to 360 days. The minimum theoretical frequency for a star observed for 28 days is 0.41 $\mu$Hz while 360 days is 0.033 $\mu$Hz. Since many stars are not observed continuously, it is not possible to calculate the minimum frequency using the inverse of the total time span of the observations. In the worst cases, i.e. stars observed only in the first and last sector of one year will have a total time span of about 360 days. However a similar star will be observed in only 2 sectors for a total time of 56 days. So for our purpose we use as minimum frequency the inverse of the number of observed sectors $N_{\text{SEC}}$ multiplied by their nominal length $L_{\text{SEC}}$. More precisely the TESS orbital period changes as function of time, hence the definition of minimum frequency need to take in account this fact, and hence for each star is:

$$f_{\min} = \frac{1}{\sum_j L_j} \quad (2)$$

where $L_j$ is the length of each sector $j$, identified from the FITS files counting all the data that have non null values in the column Time ($NN_j$) and multiplying by the cadence $dt$ coming from the header of the same file (TIMEDEL value, approx 120 seconds). Hence the observation length for a particular star is

$$\Delta T = \sum_j NN_j \cdot dt \quad (3)$$

The maximum frequency $f_{max}$ will depend on the cadence of SC observations. Considering the Nyquist criterion, the maximum observable frequency is $f_{max} = 4166\mu Hz$.

The number of frequency grid points ($n_{ls}$) has a linear dependency on the observation length:

$$n_{ls} = n_{samples} \cdot \text{INT}(\Delta T \cdot f_{max}) \quad (4)$$

where the $n_{samples}$ refers to the samples per peak (number of grid points across each peak) and we use the value of 31, $\Delta T$ is the observation length and $f_{max}$ is the maximum frequency.

One of the ways to identify the frequencies of the peaks in the periodogram is to individually fit each peak. The process is computationally expensive considering the number of peaks in the periodogram of each star. Combined with the total number of stars (which is in excess of 200,000 for the first two years alone) makes fitting individual peaks computationally prohibitive. In the context of the algorithm, the code needs to adapt to each light curve and if the light curve is flat, computational time will be lost trying to fit non-existent peaks. Therefore, a good compromise is to use a high oversampling factor which allows one to identify all the peaks clearly, whilst providing a good approximation of the frequencies of the peak without fitting them individually. We tested multiple oversampling factors > 5 on stars with visible periodic variations in the light curve, for instance eclipsing binaries. An oversampling factor of 31 samples the peak in a manner where the frequency of the middle of the peak provides a good approximation of the periodicity while also utilizing reasonable computation time.

For calculating the Lomb Scargle periodograms we used the package *LombScargle* from Astropy Collaboration et al. (2013) using the default standard normalization model which gives dimensionless power $P$.

The presence of data gaps affect the periodogram analysis with the introduction of spurious peaks in the periodogram which are not of astrophysical origin, hence we also calculate the spectral window (where we used the same frequency grid of the periodogram).

On the calculated periodograms we created the frequency list, identifying all the peaks with power above 2 times the rms of the periodogram plus the median of the periodogram (power$_{peak} > 2 \cdot \sigma_{PER}$+median$_{PER}$) and that are separated at least by their half width at half maximum (HWHM)

$$HWHM = \frac{1}{2}\frac{1}{\Delta T} \quad (5)$$

## 7 | RESULTS AND DISCUSSION

In this section we discuss some preliminary results that are obtained with `Taranga` applied to TESS year 1 & 2 data. The full description of these results is out of the scope of this article and it will be provided in a forthcoming article.

### 7.1 | SAP reconnection

The procedure of reconnecting the SAP fluxes increases the capability to study astrophysical signals that `SPOC` pipeline otherwise over correct. To show this we present in Fig. 12 the periodograms of the stars used to illustrate the reconnection procedure. It is evident that the periodogram of PDCSAP is biased and not providing the correct frequencies of the real signal. The periodograms from the original SAP and merged SAP light curves are show the frequency of the maximum peak at 0.3 $\mu$Hz and 0.2 $\mu$Hz respectively for TIC 255565783. The period of the star is 93.4 days (Watson, Henden, & Price, 2006) while the periods identified by the periodograms of original and reconnected SAP are 38.5 days and 57.9 days respectively. The reason for this is that the star was observed only for 104 days. Longer observation time will lead to better identification of the peak. However, it is clear that the frequency identified using the reconnected SAP light curve is closer to the real value compared to that identified using the original SAP.

The behaviour of AX Men is slightly different compared to HD 45555. As the original and merged SAP light curves of TIC 278140574 are not drastically different, their periodograms are also mostly similar. However, the periodograms fail to provide the full peak as the star was observed only for 84 days (the period is 100 days (Watson et al., 2006)).

### 7.2 | Period search

The identification of the highest peak in periodograms is fundamental for folding the light-curves and identifying the variability class of the objects. In APPENDIX A: we show three examples of variables stars identified phase folding their light-curves with the highest peak detected in their periodogram.

### 7.3 | Photometric results

In all the following figures, it is important to note that the missing stars in the interval of apparent magnitudes between 6 and 8 in Tmag is due to a selection criteria for SC targets by TESS team, and is not related to the present analysis.

The photometric noise of TESS in terms of rms of the normalized light-curves as function of TESS apparent magnitude is presented as a density plot in Fig. 13 . We fit the coefficients $b_0$ and $b_1$ of Eq. 6 to the rms of the light curve and the magnitude. The best fit values are reported in Table 4 .

$$rms = b_0 + b_1 e^{Tmag/b_2} \quad (6)$$

We evaluate the zero point of the PDCSAP light-curves equalling for each star its Tmag provided by the TIC catalog (Stassun et al., 2018) and the value of its median PDCSAP value:

$$Tmag = -2.5 \log_{10} M_{PDC} + z_p \quad (7)$$

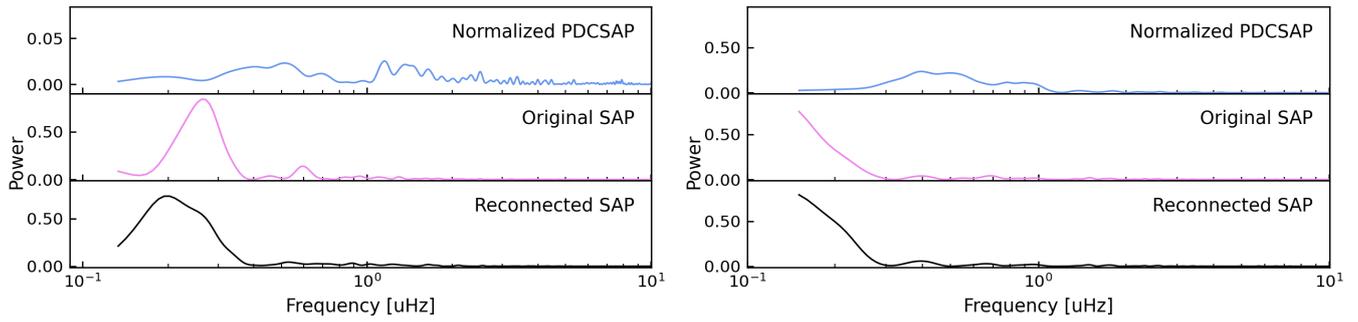

**FIGURE 12** Periodograms of the light-curves of TIC 255565783 (Fig. 10 ) on the right and TIC 278140574 (Fig. 11 ) on the left. Top: Periodogram of normalized PDCSAP Middle: Periodogram generated for the original SAP light curve. Bottom: Periodogram generated for the reconnected SAP light curve.

| coeff | value | error |
|---|---|---|
| $b_0$ | -5.04865 | 0.01660 |
| $b_1$ | 0.63124 | 0.00983 |
| $b_2$ | 8.41464 | 0.05405 |

**TABLE 4** Coefficients (Eq.6) of the fit of the lower envelope of the normalized rms as function of Tmag.

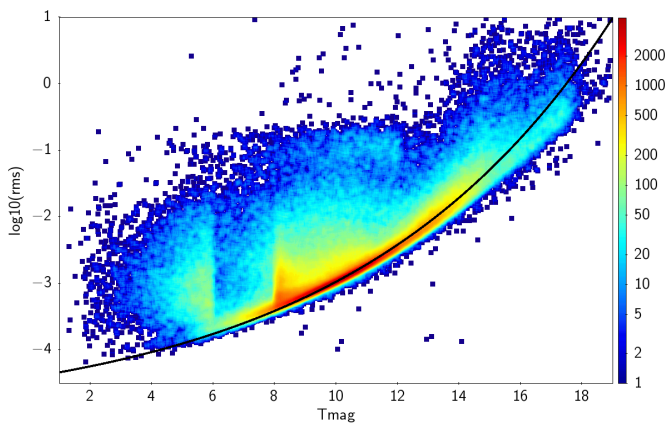

**FIGURE 13** Density plot of the normalized rms as function of Tmag. The black line represent the equation of the best fit.

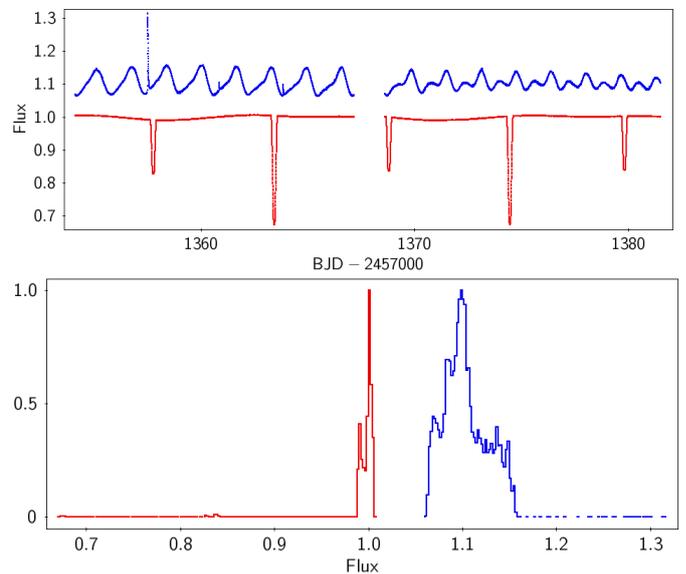

**FIGURE 14** Example light-curves and histograms for 2 stars with positive and negative skewness taken from year 1 sector 2. In Blue is TIC 229066844 (HD 217344) a RS CVn variable with some flares presents (positive skewness), in Red is TIC 179985566 (CG Scl) an Algol type eclipsing binary (negative skewness). The top panel shows the normalized light-curves of both stars, the bottom panel shows their histograms. The flux of HD 217344 (in blue) for plotting purposes is increased artificially by 0.1.

where $z_p$ is the zero point of the PDCSAP magnitudes, Tmag is the magnitude of the star, $M_{PDC}$ is the median value of the PDCSAP flux of the star. After fitting the resulting distribution with a gaussian we obtained:

$$z_p = 20.517 \pm 0.0174 \qquad \text{(PDCSAP)} \qquad (8)$$

The zero point value of the PDCSAP magnitudes we obtained is in agreement with Nardiello et al. (2019) and Handberg et al. (2021).

### 7.4 | Results from light-curves analysis

At continuation we show some figures obtained with our results that could be useful for identification of classes of stellar variability with machine learning techniques.

Since the skewness and kurtosis are measure of the shape of a distribution their values for the light-curves can be used for identification different class of variable stars. As a guiding examples for the following discussion we shown in Fig. 14 and Fig. 15 the light-curves and the histograms of some stars that have with different signs in skewness and kurtosis. In general star with negative skewness mostly corresponds to eclipsing binaries, while the ones with positives skewness are mostly stars with flares or pulsations. Stars with negatives kurtosis tends to be stars with a symmetric variation (i.e. sinusoidal variations), kurtosis near zero imply a very gaussian light-curve (no variability), positive kurtosis correspond to stars with peaked distributions and long tails (i.e. eclipsing binaries or flaring stars).

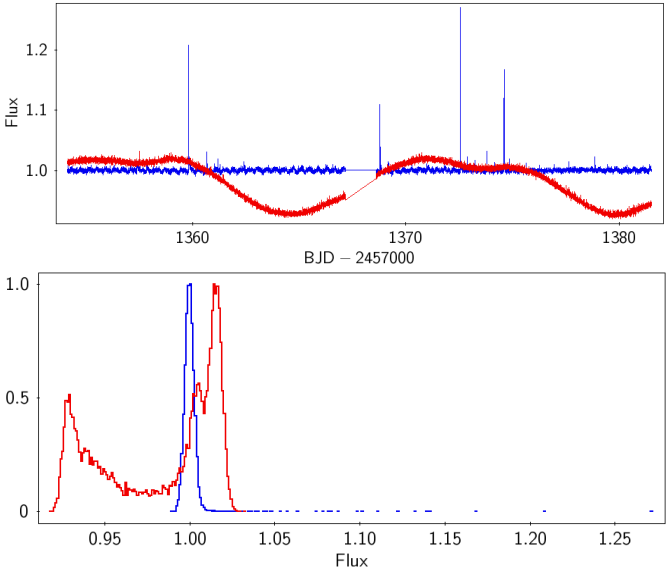

**FIGURE 15** Example light-curves and histograms for 2 stars with positive and negative kurtosis taken from year 1 sector 2. In Blue is TIC 2760232 (GJ 4360) a binary M dwarf with repeated flares (positive skewness, peaked distribution with long tail), in Red is TIC 131395566 (CD-25 16390) a rotational variable K dwarf star (negative skewness, quasi flat distribution). The top panel shows the normalized light-curves of both stars, the bottom panel shows their histograms.

In Fig. 16 we present the rms vs skewness, in the top panel the stars with negative skewness, in the bottom the stars with positive skewness). After cross check with literature we have identified some significant clusters of objects. In the top panel: the group at (-1,-0.5) is mainly composed by eclipsing binaries of type WUma, the group at (-1,+0.5) is mainly composed by eclipsing binaries of type Algol, the group at (-1.8,-1.7) is mainly composed by $\delta$ Scuti and RR Lyrae stars. In the bottom panel: the group at (-2.8,+1.0) is mainly composed by long period variables, the group at (-0.8,-0.25) is mainly composed by pulsating stars, the group at (-1.8,1.3) is mainly composed by flaring stars.

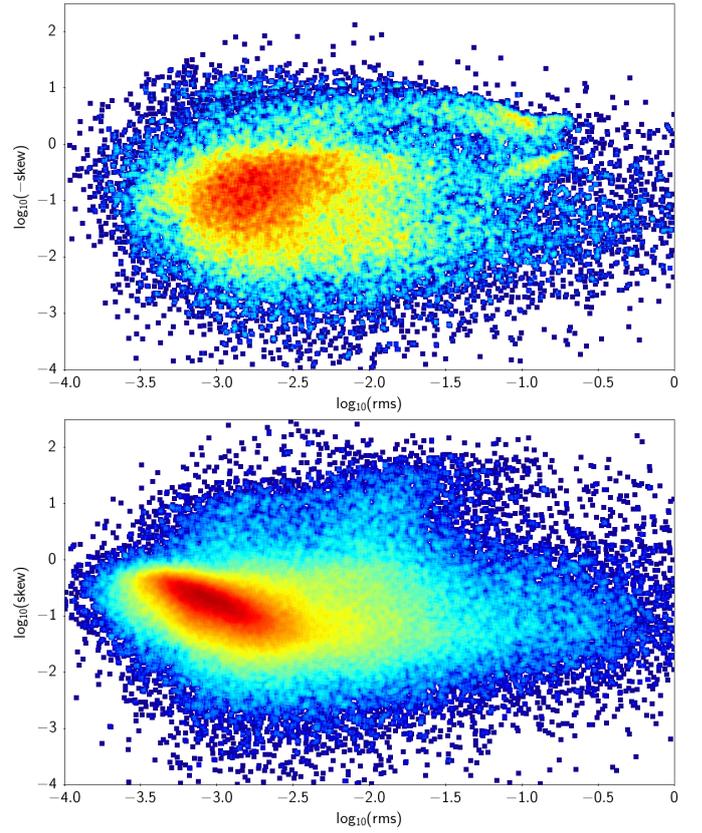

**FIGURE 16** Normalized light-curves: rms vs skewness. For a better representation of the full span of skewness we split this plot in two parts, in the top panel we show only the stars with negative skewness, in the bottom panel the stars with positive skewness. The skewness is presented in logarithmic scale, for plotting purposes in top panel the the y axis correspond to log10(-skewness). The plot represents the density of stars per spatial bin, the color scale ranges from blue (less density) to red (highest density).

## 7.5 | Background

The very good quality of the photometry of TESS allow to detect astrophysical signals also in the background, not only from the point of view of temporal evolution. Despite of the coarse angular resolution provided by the selection of the SC targets, we would like to point out how it is possible to recognize some extragalactic objects in the background measurements of TESS. In particular the luminosity function of the Large Magellanic Cloud is clearly evident in the maps of the background. For this purpose we plot the map of TESS targets centered in the LMC, and we associate to each target their background magnitude, calculated using the following formula

$$m_{\rm bkg} = -2.5 \log_{10} M_{\rm bkg} + z_p \quad (9)$$

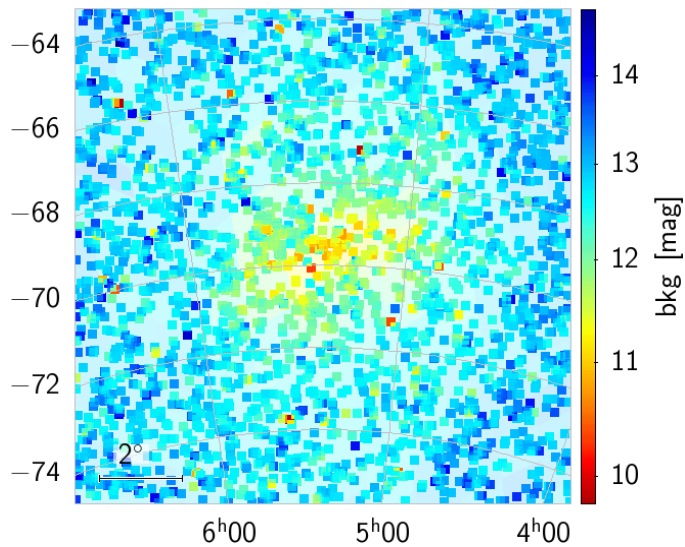

**FIGURE 17** Luminosity map of the LMC as measured from BKG median values of the targets in this part of the sky.

$m_{bkg}$ is the magnitude of the background, $M$ is the median value of the BKG flux of the target star, where $z_p$ is the zero point of the PDCSAP magnitudes previously obtained. The map of the LMC is portrayed in Fig. 17

## 8 | CONCLUSIONS

We presented a pipeline `Taranga`, for analysis of TESS SC data, and applied the pipeline to SC data of the first 2 years of the nominal mission, for a total of 232 122 stars.

The summary of the actions of the pipeline is the following:

- PDCSAP light-curves
  - merge and normalize the light-curves coming from different sectors
  - calculate the periodograms for the merged light-curves
- SAP light-curves
  - merge the light-curves coming from different sector trying to preserve the original shapes
  - calculate the periodograms for the merged light-curves

The pipeline provide as results the PDCSAP merged and normalized light-curves, the SAP reconnected light-curves, the periodograms for normalized PDCSAP and reconnected light-curves, the statistic table for light-curves and periodograms.

We plan to regularly update the analysis of SC the data as soon each year of observation of TESS become available. In particular our next data release will contain data from year 3 and 4, that will complement the ones from year 1 because they are observed in the same hemisphere. The SAP merging algorithm is presently mostly suited for AGB stars or for stars with long rotational period in the main sequence, due to the many systematics present in the SAP data we would like to include in the next data release an algorithm that includes the CBV for removing any unreal trend in the SAP data.

Finally, we note that the statistical properties of the stars observed could be useful for preparation of future photometric surveys like PLATO (Catala, 2009) and Roman Space Telescope (Montet, Yee, & Penny, 2017; Spergel et al., 2015).

## Acknowledgements

We would like to thank the reviewer for the helpful feedback on the paper. S.N wishes to acknowledge the PhD scholarship from Instituto de Investigación en Astronomia y Ciencias Planetarias, Universidad de Atacama. The following internet based resources were used in this work: NASA Astrophysics Datasystem, SIMBAD, VizieR catalog access tool operated at CDS, Strasbourg, France, VSX catalog operated by The American Association of Variable Star Observers (AAVSO), ar$\chi$iv scientific paper pre-print service operated by Cornell University

## Author contributions

S.N: Sangeetha Nandakumar
M.B: Mauro Barbieri
J.T.R: Jeremy Tregloan-Reed

S.N and M.B designed and implemented the pipeline. S.N, M.B, and J.T.R were involved in the analysis of the results and writing of the manuscript.

# APPENDIX A: ADDITIONAL MATERIAL

Here, we present some examples on the periodograms generated by `GPER` and period search.

Fig. A1 shows RR Lyrae star which was observed in only one sector for a total of 27.8 days. The periodogram shows a clear peak at 24.13 $\mu$Hz, the corresponding pulsation period is 0.4796±0.0001 days. The harmonics at 48.28 $\mu$Hz, 72.39 $\mu$Hz, and 96.54 $\mu$Hz are also identified by the pipeline.

Fig. A2 shows an eclipsing binary which was observed in sectors 25 and 26 for 48.03 days. The frequency of the highest peak is around 10.21 $\mu$Hz. The pipeline also identifies all the harmonics at 5.10 $\mu$Hz, 15.32 $\mu$Hz, 20.43 $\mu$Hz, 25.52 $\mu$Hz, 30.63 $\mu$Hz, 35.74 $\mu$Hz, and 40.85 $\mu$Hz. The period from the gaussian fit to the highest peak is around 1.13 days. VSX (Watson et al., 2006) catalog report a period of 2.2669 days. The second highest peak corresponds to this period and the gaussian fit to this peak is shown in the inset. With the fit, the period determined is 2.2674±0.0009 days.

In Fig. A3 shows a M dwarf with short orbital period flares. The periodogram shows a clear peak at 8.9 $\mu$Hz corresponding to a period of 1.3 days.

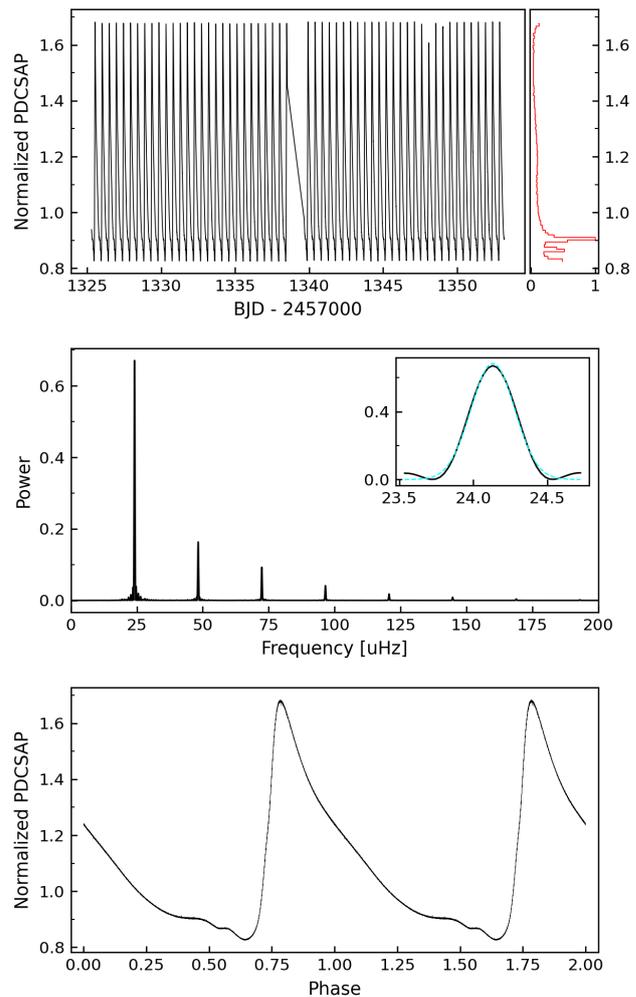

**FIGURE A1** TIC 126910093: V Ind, an A5 star of RR Lyrae type, observed in 1 sector during year 1. Top: The normalized light curve along with the histogram of the normalized flux shown in red. Middle: The lomb scargle periodogram with the inset showing the zoom of the highest peak with gaussian fit. The pulsation period determined from the gaussian fit to the highest peak in the periodogram is 0.4796±0.0001 days. Bottom: The light curve folded with the determined period.

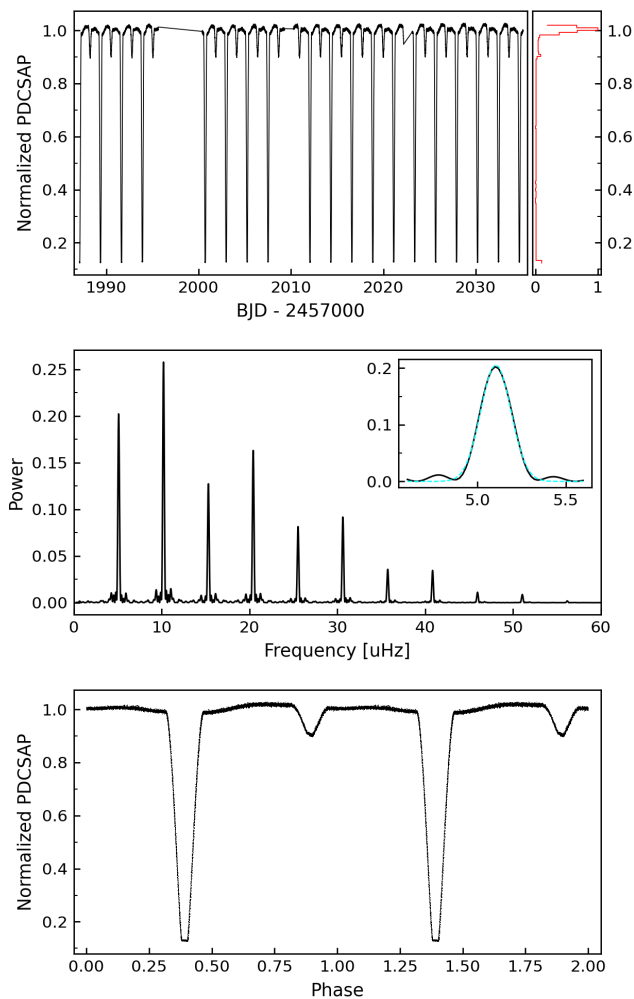

**FIGURE A2** TIC 9473243: TU Her, a F5 an Algol type eclipsing binary. Top: Normalized light curve with the histogram shown in red Middle: The periodogram and the inset showing the gaussian fit to the highest peak. The orbital period from literature is P=2.2669385 (Watson et al., 2006). Orbital period determined from periodogram P=2.2674±0.0009 (frequency of the second highest peak) Bottom: Light curve folded with the determined period

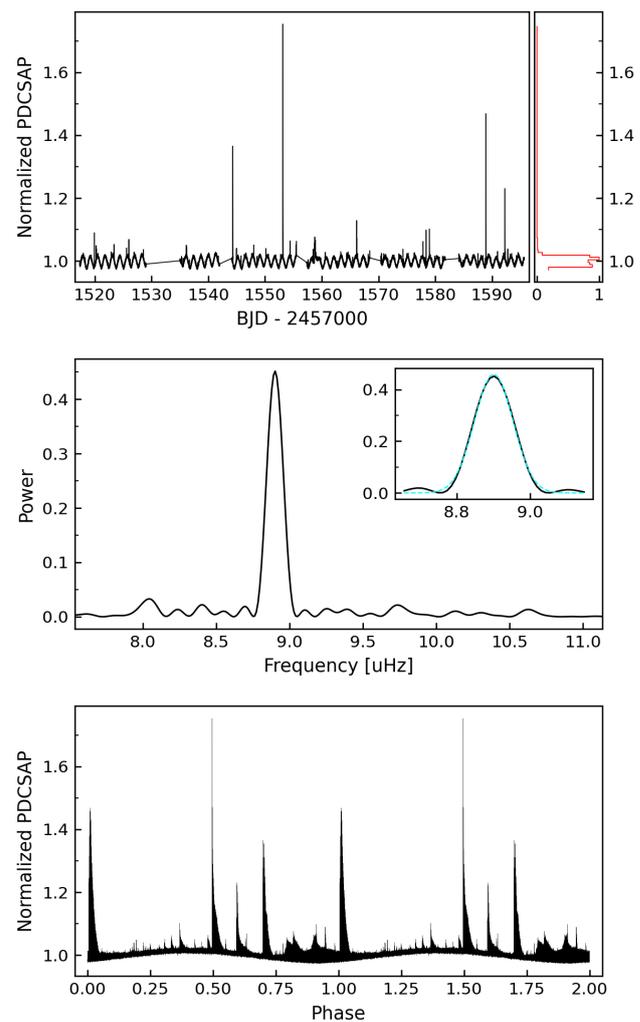

**FIGURE A3** TIC 46029665: A M4 dwarf, known as X rays source also. Top: Normalized light curve with the histogram shown in red. Middle: The periodogram and the inset showing the gaussian fit to the highest peak. The period determined from the gaussian fit to the highest peak in the periodogram is 1.3002±0.0002 days Bottom: Light curve folded with the above period